\documentclass[%
 reprint,
 superscriptaddress,
 amsmath, amssymb,
 prb,
 citeautoscript, 
]{revtex4-1}

\usepackage{graphicx}
\usepackage{dcolumn}
\usepackage{bm}
\usepackage{placeins}
\usepackage{algorithm}
\usepackage{algpseudocode}
\usepackage{color,soul}
\usepackage[version=3]{mhchem}

\begin{document}

\title{Efficient periodic density functional theory calculations of charged molecules and surfaces using Coulomb kernel truncation}

\author{Sudarshan Vijay}
\email{sudarshan.vijay@vasp.at}
\email{sudarshan.vijay@iitb.ac.in}
\affiliation{VASP Software GmbH, Berggasse 21, 1090 Vienna, Austria}
\address{Present address: Department of Chemical Engineering, Indian Institute of Technology Bombay, Powai, Mumbai, Maharashtra 400076 India}
\author{Martin Schlipf}
\author{Henrique Miranda}
\author{Ferenc Karsai}
\author{Merzuk Kaltak}
\author{Martijn Marsman}
\affiliation{VASP Software GmbH, Berggasse 21, 1090 Vienna, Austria}
\author{Georg Kresse}
\email{georg.kresse@univie.ac.at}
\affiliation{VASP Software GmbH, Berggasse 21, 1090 Vienna, Austria}
\affiliation{Faculty of Physics and Center for Computational Materials Science, University of Vienna, Kolingasse 14-16, A-1090 Vienna, Austria}%

\date{\today}

\begin{abstract}
    Density functional theory (DFT) calculations of charged molecules and surfaces are critical to applications in electro-catalysis, energy materials and related fields of materials science.
    DFT implementations such as the Vienna \textit{ab-initio} Simulation Package (VASP) compute the electrostatic potential under 3D periodic boundary conditions, necessitating charge neutrality.
    In this work, we implement 0D and 2D periodic boundary conditions to facilitate DFT calculations of charged molecules and surfaces respectively.
    We implement these boundary conditions using the Coulomb kernel truncation method.
    Our implementation computes the potential under 0D and 2D boundary conditions by selectively subtracting unwanted long-range interactions in the potential computed under 3D boundary conditions.
    By combining the Coulomb kernel truncation method with a computationally efficient padding approach, we remove nonphysical potentials from vacuum in 0D and 2D systems.
    To illustrate the computational efficiency of our method, we perform large supercell calculations of the formation energy of a charged chlorine defect on a sodium chloride (001) surface and perform long time-scale molecular dynamics simulations on a stepped gold (211) $\vline$ water electrode-electrolyte interface.
\end{abstract}

\maketitle
\section{Introduction}
\label{sec:introduction}

Efficient density functional theory calculations of charged molecules and surfaces are crucial to predict novel electro catalysts\cite{shin_importance_2022,lindgren_electrochemistry_2022}, reliable battery materials \cite{Spotte-Smith2022a,kumar_crystal_2014} and stable defects in surfaces of semi-conductors. \cite{Komsa2014,Freysoldt2014}
Commonly used DFT implementations such as the Vienna \textit{ab-initio} Simulation Package (VASP)\cite{Kresse1996} rely on periodic boundary conditions in all dimensions (3D boundary conditions). 
While this assumption of periodicity allows for computationally efficient fast-Fourier transforms (FFT) to solve the Poisson equation (to compute the potential from a charge density), it imposes restrictions on the dimensionality of materials that can be computed without the influence of spurious electrostatic interactions from periodic replicas.

There are two constraints with using 3D boundary conditions to solve the Poisson equation. First, we implicitly assume that \emph{both} the charge density and the resulting potential are periodic in all dimensions.\cite{ewald_berechnung_1921,noauthor_simulation_1980}
Bulk materials satisfy this criterion and are hence trivially treated by this method.
Conversely, systems with lower dimensionality, such as 2D and quasi-2D systems (surfaces), 1D systems (nano-wires) and 0D systems (molecules) are partially or fully non-periodic and require large cells to remove spurious electrostatic interactions between unwanted periodic replicas.\cite{Bengtsson1999,neugebauer_adsorbate-substrate_1992,Sohier2017,Makov1994,jarvis_supercell_1997}
Second, this reciprocal space method requires charge-neutrality, i.e.\ the electronic and nuclear charges must sum to zero.\cite{ewald_berechnung_1921}
Total energies of systems that are not charge neutral are divergent and hence of limited use.\cite{noauthor_simulation_1980}

The first challenge of lower-dimensional periodicity has been approached by Coulomb kernel truncation techniques\cite{jarvis_supercell_1997,Rozzi,Sohier2017}.
These methods adapt the Poisson equation to lower dimensionality.
However, they require padded cells at least 27$\times$  larger than the conventional cells for 0D systems and 2$\times$ larger for 2D systems.\cite{jarvis_supercell_1997} 
Padding makes this procedure computationally expensive, as FFTs scale with the supercell dimension $N$ as $N\log N$. 
This expense precludes large supercell calculations and long time-scale \textit{ab-initio} molecular dynamics simulations.
Recent implementations\cite{Rozzi,Sohier2017} circumvent the use of padding by requiring a large vacuum along the non-periodic dimension.
However, this choice leads to artifacts in the potential at the edges of the cell and requires convergence with respect to both vacuum dimension and choice of kernel truncation.
Alternative approaches such as modifying the basis functions to be non-periodic in certain dimensions prevent the need for padding, but require specialized treatments such as using the full-potential linearized augmented plane-wave (FLAPW) method.\cite{mokrousov_full-potential_2005}

The second challenge of charge neutrality has been addressed by applying a compensating homogeneous background charge over the entire cell.
Alternatively, a compensating charge is added to regions of the cell where there is a vanishing charge density, such as in the case of implicit solvent methods.\cite{Andreussi2014,lindgren_electrochemistry_2022,Mathew2014,Mathew2014a,Sundararaman2017a} 
While this compensating charge may be physically motivated in certain applications, it often introduces additional artifacts into the computation of the total energy which must be treated \textit{post facto}.\cite{hub_quantifying_2014,Gauthier2019c}

In this work, we solve both these challenges by implementing a method for applying established Coulomb kernel truncation methods for systems with 0D and 2D boundary conditions \emph{without} the need for increased FFT grids.
Our approach allows for computing large systems efficiently and is suitable for performing long-timescale molecular dynamics simulations.
With this method, we treat charged 2D periodic surfaces, effectively removing the requirement of charge neutrality.
Throughout this manuscript, we detail specifics of our method as implemented in VASP.
In Section \ref{sec:boundary_conditions}, we review Coulomb kernel truncation methods to alter 3D boundary conditions to 0D and 2D boundary conditions.
In Section \ref{sec:coarsening}, we describe our approach of Coulomb truncation without padding to a large supercell.
In Section \ref{sec:energies}, we compare the variation of the total energy between 3D, 2D and 0D boundary conditions.
In Section \ref{sec:applications}, we illustrate the computational advantage of our approach by calculating the dependence of the formation energy on the concentration of charged vacancy defects of sodium chloride (NaCl).
In addition, we perform charged \textit{ab-initio} molecular dynamics simulations of a stepped  Au(211) $\vline$ water interface over long timescales.
\section{Coulomb kernel truncation and charged systems}
\label{sec:boundary_conditions}

In this section, we compute the potential under 0D and 2D boundary conditions by truncating the Coulomb kernel.\cite{Rozzi,Sohier2017,jarvis_supercell_1997}
We highlight modifications to the current 3D implementation of VASP and outline the procedure to compute the potential of charged surfaces and molecules without any \textit{post hoc} corrections.
As an illustration, we discuss differences in the computed potential through a representative example of charge-neutral and positively-charged MoS$_2$ with and without applying the Coulomb kernel truncation method.

\subsection{Coulomb kernels for 3D, 2D and 0D boundary conditions}
\label{subsec:kernels}
We determine the potential ($V$) from a given charge distribution ($\rho$) using the Poisson equation
\begin{equation}
    \label{eq:poisson}
    V(\mathbf{r}) = \int \frac{\rho(\mathbf{r}^\prime)}{\left | \mathbf{r} - \mathbf{r}^\prime \right|} d\mathbf{r}^\prime,
\end{equation}
\noindent where $\mathbf{r}$ denotes a position in real space.
In a fully periodic calculation, the integral in Eq.\ \eqref{eq:poisson} is performed over all repeated cells and the periodic charge density $\varrho(\mathbf{r})$ as
\begin{equation}
    \label{eq:poisson-periodic}
    V(\mathbf{r}) =  \int \frac{\varrho(\mathbf{r}^\prime)}{\left | \mathbf{r} - \mathbf{r}^\prime \right|} d\mathbf{r}^\prime.
\end{equation}
\noindent However, since the potential is also cell-periodic, $\mathbf{r}$ can be restricted to the primitive cell.
Fourier transforming Eq.~\eqref{eq:poisson-periodic} to reciprocal space, we get

\begin{equation}
    \label{eq:poisson-periodic-reciprocal}
    V(\mathbf{g}) = \frac{4\pi}{\mathrm{g}^2} \varrho(\mathbf{g}),
\end{equation}
\noindent where $\mathbf{g}$ is a reciprocal lattice vector. 
We use non-bold versions of the symbol (for example, $\mathrm{g}$) to indicate the norm of the vector of the symbol (for example $\mathbf{g}$).
We introduce the Coulomb kernel $v_{\text{3D}}(\mathbf g) = 4\pi/\mathrm{g}^2$ to express Eq.~\eqref{eq:poisson-periodic-reciprocal} as
\begin{equation}
   \label{eq:coulomb-kernel}
   V(\mathbf{g}) = v_{\text{3D}}(\mathbf{g}) \varrho(\mathbf{g}).
\end{equation}
Physically, applying this kernel to $\varrho$ generates a potential where each point feels the entire $1/r$ Coulomb interaction from every other point in space. 

We follow Ref.~\onlinecite{Rozzi} and implement 0D and 2D periodic boundary conditions by restricting the Coulomb interaction to a certain region ($\mathcal{R}$)
\begin{equation}
    \label{eq:truncated-kernel-base}
    v(\mathbf{r}) =
    \begin{cases}
        1/r \quad & r \in \mathcal{R} \\
        0 \quad & r \notin \mathcal{R}.
    \end{cases}
\end{equation}
For 0D boundary conditions, we truncate the kernel in real space to a sphere\cite{jarvis_supercell_1997}
\begin{equation}
    v(\mathbf{r}) = \frac{1}{r} \theta(R_\mathrm{c} - r),
\end{equation}
where $R_\mathrm{c}$ is the radius of the sphere and $\theta$ is the Heaviside function. Similarly, $\mathcal{R}$ is a cuboid for 2D boundary conditions with the real-space kernel\cite{Rozzi,Sohier2017}
\begin{equation}
    v(\mathbf{r}) = \frac{1}{r}\theta(R_\mathrm{c} - r_z).
\end{equation}
Here, $R_\mathrm{c}$ is the dimension of the cuboid in the direction of the surface normal and $r_z$ is the position along the surface normal.
Solving the Poisson equation in reciprocal space, we require the Fourier-transformed kernels Ref.\ \onlinecite{Rozzi,jarvis_supercell_1997}  for 0D boundary conditions
\begin{equation}
    \label{eq:kernel-0D}
    v_{\text{0D}}(\mathbf{g}) = 
    \begin{cases}
        4\pi/\mathrm{g}^2 \left( 1 - \cos\left(\mathrm{g}R_\mathrm{c}\right) \right) & \quad \mathrm{g}\neq 0 \\
        2\pi R_\mathrm{c}^2 & \quad \mathrm{g} = 0,
    \end{cases}
\end{equation}
and for 2D boundary conditions in Ref.\ \onlinecite{Rozzi}
\begin{widetext}
\begin{equation}
    \label{eq:kernel-2D}
    v_{\text{2D}}(\mathbf{g}) = 
    \begin{cases}
        4\pi / \mathrm{g}^2  \left[ 1 - e^{-\mathrm{g}_{\parallel}R_c} \left ( \cos(\mathrm{g}_\perp R_\mathrm{c}) + \mathrm{g}_\perp / \mathrm{g}_{\parallel} \sin(\mathrm{g}_{\perp}R_\mathrm{c}) \right ) \right] \quad & \mathrm{g} \neq 0 \\
        4\pi / \mathrm{g}_\perp^2\left[ 1 - \cos(\mathrm{g}_\perp R_\mathrm{c}) - \mathrm{g}_{\perp}R_\mathrm{c} \sin(\mathrm{g}_\perp R_\mathrm{c}) \right] \quad & \mathrm{g}_{\parallel} = 0 \\
        -2\pi R_\mathrm{c}^2 \quad & \mathrm{g}=0,
    \end{cases}
\end{equation}
\end{widetext}
where $\mathrm{g}_{\parallel}$ is the norm of the $\mathbf{g}$ vector parallel to the surface and $\mathrm{g}_{\perp}$ is the projection of the $\mathbf{g}$ vector on the surface normal.
Note that for the specific choice of $R_\mathrm{c} = L_z/2$, the $\sin$ terms of Eq. (\ref{eq:kernel-2D}) go to zero and there is no divergence at $\mathrm{g}_\parallel \to 0$.\cite{ismail-beigi_truncation_2006}

\subsection{Constructing padded supercells}

In the previous section, we presented Coulomb kernels for 0D and 2D dimensionalities which depend on a truncation parameter $R_\mathrm{c}$.
In this section, we discuss how $R_\mathrm{c}$ is chosen in practice for 0D and 2D dimensionalities.

\begin{figure}[!htb]
    \centering
    \includegraphics[width=\linewidth]{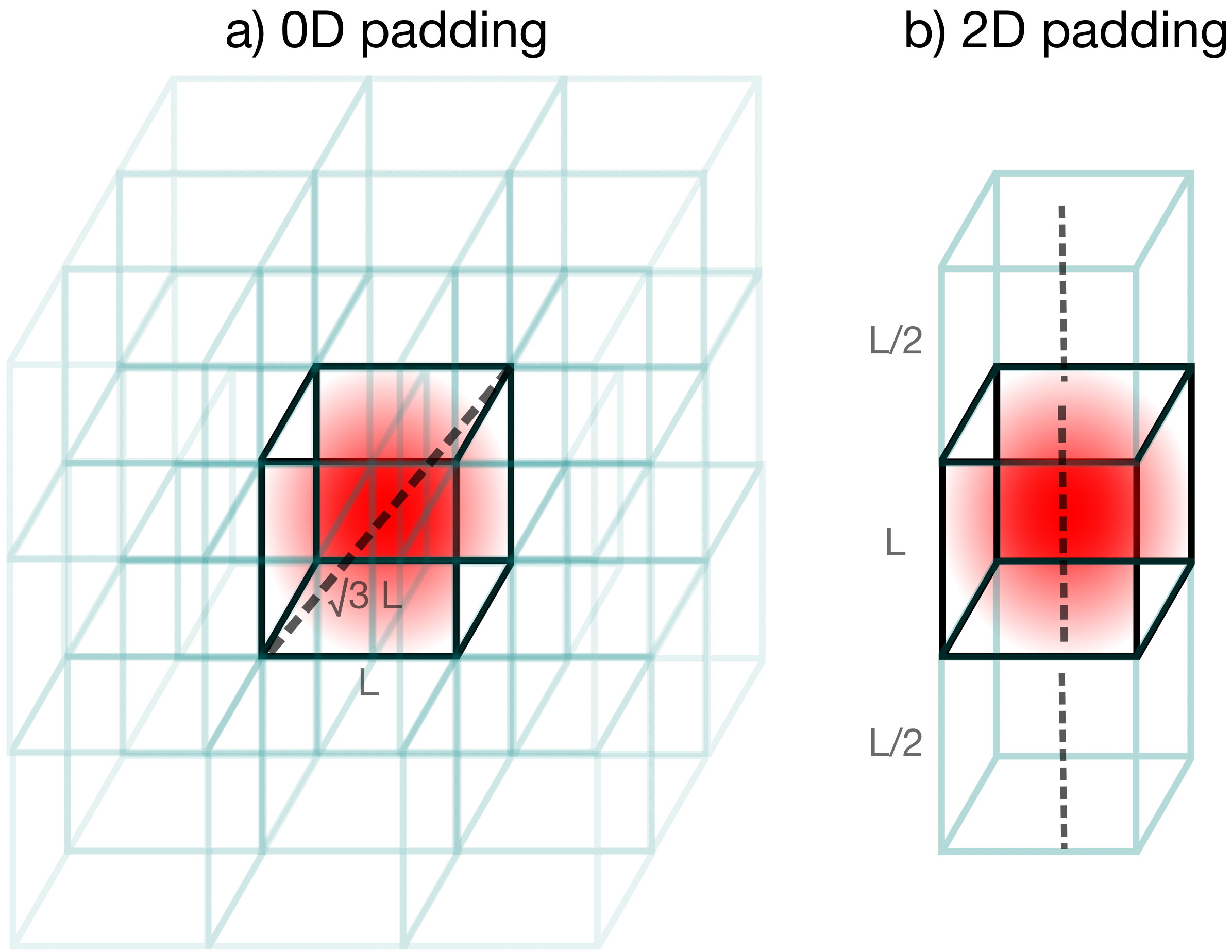}
    \caption{Schematics showing the construction of padded supercells for a) 0D boundary conditions and b) 2D boundary conditions; the red sphere denotes the original charge density and the blue cells represent the padded cells.}
    \label{fig:schematic-conventional-padding}
\end{figure}

$R_\mathrm{c}$ must be sufficiently large such that the non-periodic charge density, $\rho$, does not interact with its periodic counterpart across nonperiodic boundaries.\cite{Sohier2017,Rozzi,hine_electrostatic_2011}
We prevent this spurious interaction by extending the region $\mathcal{R}$ in Eq.\ \eqref{eq:truncated-kernel-base} to the edges of the cell and computing the Poisson equation in a padded supercell.

Fig.\ \ref{fig:schematic-conventional-padding} shows a schematic of the padded supercell used for 0D and 2D systems.
For 0D boundary conditions, $\rho$ (shown in red in Fig.~\ref{fig:schematic-conventional-padding}~a) must not interact with any of its periodic replicas.
Padding the cell in every dimension (blue cells in Fig.~\ref{fig:schematic-conventional-padding}~a) fulfills this requirement but leads to a supercell that is $3^3=27$ times larger than the otherwise considered cell.\cite{Rozzi}
For 2D boundary conditions, $\rho$ must not interact with its periodic replicas in the direction of the surface normal; requiring padding of $2\times$ the norm of the lattice vector along one dimension (blue cells in Fig.\ \ref{fig:schematic-conventional-padding}~b).\cite{Sohier2017,Rozzi}

These padded cells lead to a potentially large computational cost associated with applying 0D and 2D boundary conditions in practice.
We defer discussion of our method for applying 0D and 2D boundary conditions without the need for extended FFT grids to Section \ref{sec:coarsening}.

\subsection{Comparing 2D and 3D potential for charge neutral and charged systems}

In this section, we compare the potential computed using 3D and 2D boundary conditions as discussed in Section~\ref{subsec:kernels}.
We illustrate the differences in 2D and 3D boundary conditions by computing the potential of a charge-neutral and positively charged single layer of MoS$_2$.

\begin{figure}[!htb]
    \centering
    \includegraphics[width=\linewidth]{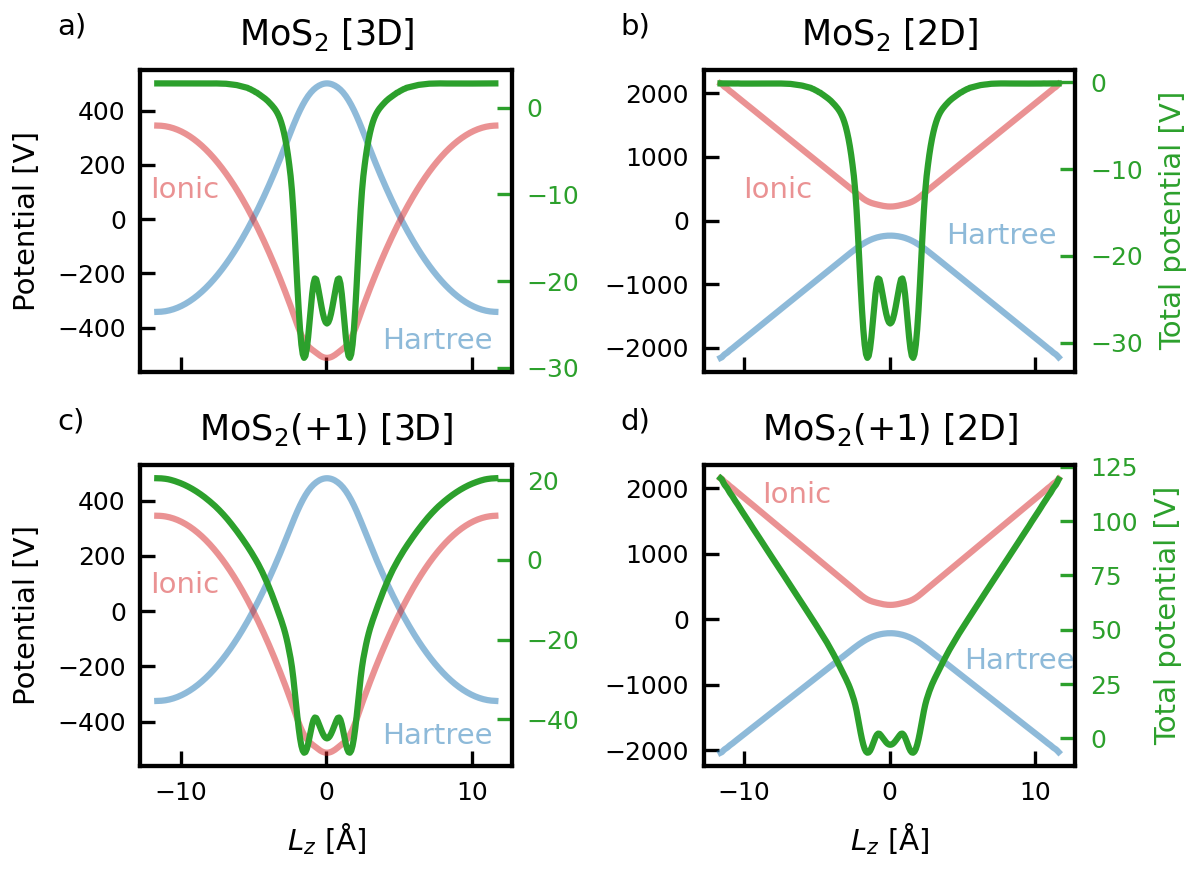}
    \caption{Comparison of the Hartree, ionic and total (sum of Hartree, ionic and exchange-correlation) potential for a) neutral MoS$_2$ with 3D boundary conditions b) neutral MoS$_2$ with 2D boundary conditions c) positively charged MoS$_2$ with 3D boundary conditions d) positively charged MoS$_2$ under 2D boundary conditions. Tick labels for the total potential (green) are on the right.}
    \label{fig:comparison-potential}
\end{figure}

Fig.\ \ref{fig:comparison-potential} shows the electron-electron Hartree potential (in blue), ion-electron (in red), and the total potential, i.e.\ sum of Hartree, ionic and exchange-correlation potentials (in green) for a single layer of charge neutral and charged MoS$_2$ under 3D and 2D boundary conditions.
The Hartree potential is obtained through Eq.~\eqref{eq:poisson-periodic-reciprocal} where $\varrho$ is the valence charge density,
\begin{equation}
    V_{\mathrm{Hartree}}(\mathbf{g}) = v(\mathbf{g}) \varrho(\mathbf{g}).
\end{equation}
Note that $v(\mathbf{g})$ is replaced by $v_{2\mathrm{D}}(\mathbf{g})$ or $v_{0\mathrm{D}}(\mathbf{g})$ to compute the Hartree potential for 2D or 0D systems respectively.
As in the formulation of Ref.~\onlinecite{ihm_momentum-space_nodate}, the ionic potential is determined by setting $\varrho$ to the ion charge as if they were point charges and adding the local contributions of the pseudopotential for each atom $i$, $U_{\mathrm{ps,i}}$,
\begin{equation}
    V_{\mathrm{ionic}} = \sum_{i} \left( v(\mathbf{g}) \frac{Z_i}{\Omega} + U_{\mathrm{ps},i} \right),
\end{equation}
where $Z_i$ is the valence charge for each atom $i$ and $\Omega$ is the volume. Note that we do not truncate the local pseudopotential as the corresponding charge density ($q_i/\Omega + \varrho_{\mathrm{ps},i}$, where $q_i$ is the ion charge corresponding to the core charge) is short-ranged and does not influence the potential beyond a few Angstroms.\cite{Sohier2017}

Under 3D boundary conditions, the Hartree and ionic potential (blue and red lines in Figure \ref{fig:comparison-potential}a respectively) show parabolic behavior in vacuum, i.e.\ they have $L_z^2$ dependence far away from the material.
Conversely, the same system computed with 2D boundary conditions shows linear behavior in $L_z$ (blue and red lines in Figure \ref{fig:comparison-potential}b) of the potential in vacuum.
When summed together for a neutral system (green lines in Figure \ref{fig:comparison-potential}a,b), the parabolic and linear dependence on $L_z$ is removed from 3D and 2D boundary conditions respectively.
The only difference in the two potentials is that the total potential decays to zero far away from the material under 2D boundary conditions while it decays to an arbitrary value under 3D boundary conditions (value of the green lines at $L_z = \pm 10$).

This parabolic behavior of the Hartree and ionic potential under 3D boundary conditions stems from the convention of setting the $\mathrm{g}=0$ term of the 3D Coulomb kernel to zero.\cite{ihm_momentum-space_nodate,Dabo} This choice implicitly represents solving the Poisson equation with the charge density set to the sum of $\varrho$ and a compensating homogeneous background charge with magnitude equal to the negative of the magnitude of $\varrho$. For charge-neutral systems, the homogeneous background charge contribution for the Hartree and ionic potentials cancel out (as they are of equal and opposite magnitudes) and the resulting total potential is free from parabolic behavior.

Conversely, the linear behavior of the Hartree and ionic potential under 2D boundary conditions is the appropriate behavior of a charged 2D sheet such as MoS$_2$.
We expect a charged sheet of charge $q$ and area $A$ to have a potential of $2\pi z \, q/A $ where $z$ is the distance along the surface normal.\cite{jackson_classical_1999}
This linear behavior is seen for both the Hartree and ionic potentials in Figure \ref{fig:comparison-potential}b and no homogeneous background charge is applied in either case.
For charge-neutral systems, this linear dependence is canceled out when summing together the Hartree and ionic potentials, and hence the total potential decays to the reference value of zero in vacuum.

Unlike charge-neutral systems, charged systems do not exhibit equivalent behavior of the total potential for 3D and 2D boundary conditions.
In contrast to charge neutral systems, the total potential (blue line in Fig.\ \ref{fig:comparison-potential}c,d) exhibits parabolic dependence on $L_{z}$.
This parabolic dependence is caused by the lack of cancellation of the potential from the homogeneous background charge when summing the Hartree and ionic potential for a charged system.\cite{Dabo}
There is no such parabolic dependence for the potential computed under 2D boundary conditions. Instead, a linear dependence of the potential commensurate with a charged 2D sheet is observed for the Hartree, ionic, and total potentials.

In summary, Fig.\ \ref{fig:comparison-potential} shows that applying 2D boundary conditions for 2D materials allows us to compute the potential for charged systems.
Applying 3D boundary conditions to charged 2D materials generates artifacts in the total potential due to the presence of a homogeneous background charge.
\section{Efficient padding for surfaces and molecules}
\label{sec:coarsening}

In Section \ref{sec:boundary_conditions}, we showed that applying the appropriate kernel and padding a cell with vacuum allows us to compute the potential for 0D and 2D boundary conditions.
A key challenge with applying this method is the need for computationally expensive padded cells.
In this section, we discuss the computational bottlenecks with the currently used \textit{padded supercell} method and present our \textit{coarsen before padding} method, which performs kernel truncation without the need for enlarged FFT grids.
We compare the two approaches against the \textit{no-padding} method, which performs truncation without padding.

\begin{figure*}[!htb]
    \centering
    \includegraphics[width=0.6\linewidth]{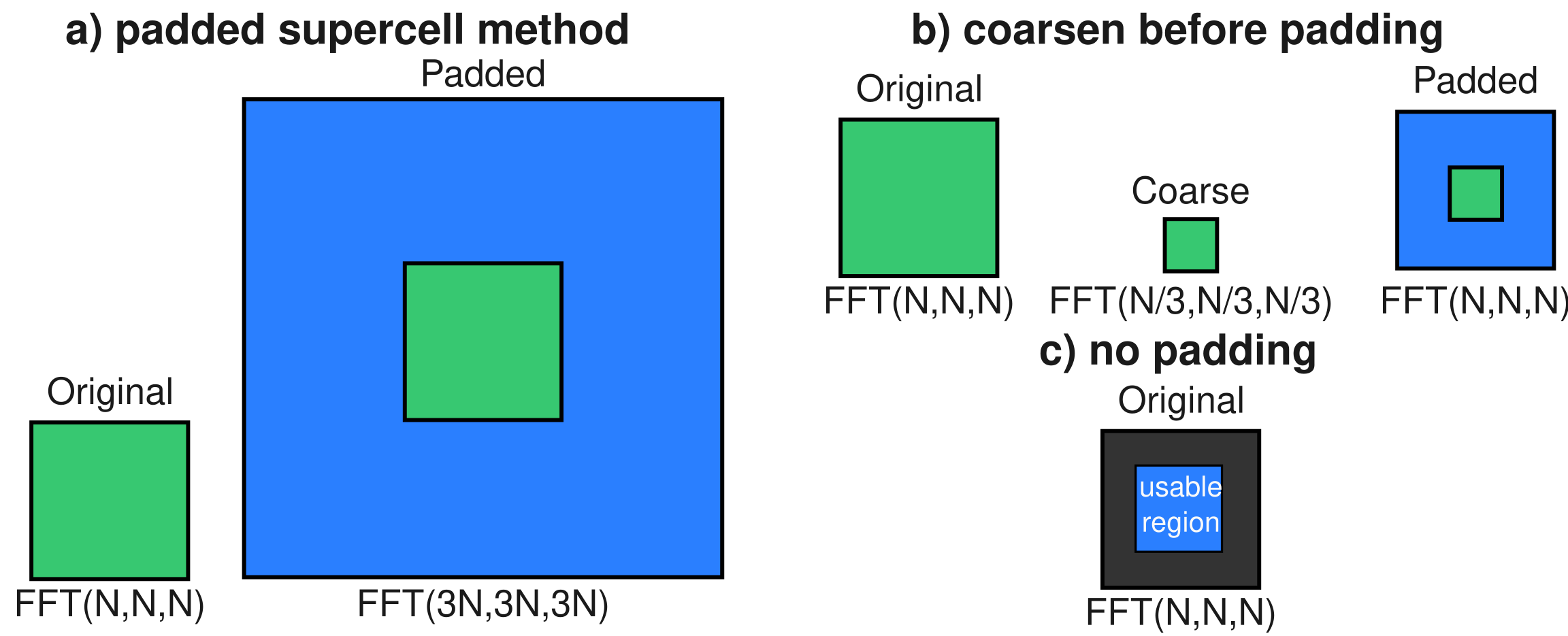}
    \caption{Schematic of a) padding supercell method using a larger grid b) coarsen before padding method that coarsens the original grid before padding and c) no-padding method which performs truncation without padding for a 0D system; the usable region shows where the potential obtained is the correct one.}
    \label{fig:coarsen-schematic}
\end{figure*}

\subsection{Padded supercell method}

\begin{algorithm}[H]
\caption{Padded supercell method}\label{alg:original-padding}
\begin{algorithmic}[1]
\State $\rho(\mathbf{r}) = \mathrm{Pad}\left[\varrho(\mathbf{r})\right]$
\State $\rho(\mathbf{g}) = \mathcal{F}[\rho(\mathbf{r})]$
\State $V_{\mathrm{pad}}(\mathbf{g}) = v(\mathbf{g})\rho(\mathbf{g})$
\State $V_\mathrm{pad}(\mathbf{r}) = \mathcal{F}^{-1} V_{\mathrm{pad}}(\mathbf{g})$
\State $V(\mathbf{r}) = \mathrm{Unpad}\left[V_{\mathrm{pad}}(\mathbf{r})\right]$
\end{algorithmic}
\end{algorithm}

Algorithm~\ref{alg:original-padding} illustrates the padded supercell approach for determining the potential. 
First, the periodic charge density $\varrho$ is padded to create a non-periodic charge density $\rho$ (Line~1, Algorithm~\ref{alg:original-padding}).
An FFT (denoted as $\mathcal{F}$) is performed on this padded cell to generate $\rho(\mathbf{g})$ (Line~2, Algorithm~\ref{alg:original-padding}).
As FFTs scale $\mathcal{O}\left(N\log N\right)$, this FFT is the largest computational expense of this method (schematically shown in Fig.~\ref{fig:coarsen-schematic}a).
For 0D systems, the padded cell is a grid of 3N$\times$3N$\times$3N (in blue) where N is the number of points in the cell. This padded grid is 27 times as large as the original N$\times$N$\times$N grid (in green).

Following this expensive FFT, the padded potential $V_{\mathrm{pad}}(\mathbf{g})$ is computed using Eq.~\eqref{eq:coulomb-kernel} with the appropriate kernel (Line~3, Algorithm~\ref{alg:original-padding}).
An inverse Fourier transform (denoted as $\mathcal{F}^{-1}$) converts it to real space, $V_{\mathrm{pad}}(\mathbf{r})$ (Line~4, Algorithm~\ref{alg:original-padding}).
This inverse FFT is equally expensive as the FFT used to compute $\rho(\mathbf{g})$ (Line~2, Algorithm~\ref{alg:original-padding}) and doubles the computational cost. 

Finally, $V_{\mathrm{pad}}(\mathbf{r})$ is condensed in real space to generate $V(\mathbf{r})$, i.e.\ grid points corresponding to the points in the original grid are retained while those corresponding to the padded grid are removed (Line~5, Algorithm~\ref{alg:original-padding}). 

\subsection{Coarsen before padding method}

\begin{algorithm}[H]
\caption{Coarsening before padding method}\label{alg:coarsening-padding}
\begin{algorithmic}[1]
\State $\varrho(\mathbf{g}) = \mathcal{F}[\varrho(\mathbf{r})]$
\State $V_{\mathrm{periodic}}(\mathbf{g}) = 4\pi/\mathrm{g}^2 \varrho(\mathbf{g})$
\If{$\mathbf{g} = \mathbf{g}_{\mathrm{coarse}}$ }
    \State $\rho_{\mathrm{coarse}}(\mathbf{g}_{\mathrm{coarse}}) = \varrho(\mathbf{g}_{\mathrm{coarse}})$
\EndIf
\State $V_{\mathrm{periodic,coarse}}(\mathbf{g}_{\mathrm{coarse}}) = 4\pi / \mathrm{g}^2_{\mathrm{coarse}} \rho_{\mathrm{coarse}}(\mathbf{g}_{\mathrm{coarse}})$
\State $\rho_{\mathrm{coarse}}(\mathbf{r}_{\mathrm{coarse}}) = \mathcal{F}^{-1}\left[\rho_{\mathrm{coarse}}(\mathbf{g}_{\mathrm{coarse}})\right]$
\State Apply Algorithm \ref{alg:original-padding} to $\rho_{\mathrm{coarse}}$ to obtain $V_{\mathrm{coarse}}(\mathbf{r}_{\mathrm{coarse}})$
\State $V_{\mathrm{coarse}}(\mathbf{g}_{\mathrm{coarse}}) = \mathcal{F}[V_{\mathrm{coarse}}(\mathbf{r}_{\mathrm{coarse}})]$
\State $V(\mathbf{g}) = V_{\mathrm{periodic}}(\mathbf{g}) - V_{\mathrm{periodic,coarse}}(\mathbf{g}_{\mathrm{coarse}}) + V_{\mathrm{coarse}} (\mathbf{g}_{\mathrm{coarse}})$
\State $V(\mathbf{r}) = \mathcal{F}^{-1}[V(\mathbf{g})]$
\end{algorithmic}
\end{algorithm}

Instead of computing the entire potential with 0D or 2D boundary conditions, our \textit{coarsen before padding} method selectively subtracts away truncated components from a 3D periodic potential to achieve 0D or 2D boundary conditions. The sequence of steps to achieve this selective computation are:
\begin{enumerate}
    \item Fourier transform the 3D periodic charge density and compute the periodic potential, $V_{\mathrm{periodic}}$ (Lines~1 and 2 of Algorithm~\ref{alg:coarsening-padding}).
    We will selectively subtract the truncated interactions from $V_{\mathrm{periodic}}$ at the end of the algorithm.
    \item Generate a coarse grid consisting of N/3$\times$N/3$\times$N/3 (see Fig.~\ref{fig:coarsen-schematic}b) for 0D boundary conditions and N$\times$N$\times$N/2 for 2D boundary conditions, where N$\times$N$\times$N is the dimension of the original cell.
    Note that an unequal number of grid points is allowed by this method.
    Transfer elements of $\varrho$ corresponding to this coarse grid to generate coarse charge density, $\rho_{\mathrm{coarse}}$ (Lines~3--5 of Algorithm~\ref{alg:coarsening-padding}).
    \item Compute the 3D periodic potential of $\rho_{\mathrm{coarse}}$ to generate $V_{\mathrm{periodic,coarse}}$ (Line~6 of Algorithm~\ref{alg:coarsening-padding}).
    This potential contains unwanted long-range periodic interactions included in $V_{\mathrm{periodic}}$ and will be subtracted from it.
    \item Perform the conventional padding procedure as seen in Algorithm~\ref{alg:original-padding} on $\rho_{\mathrm{coarse}}$ to generate $V_{\mathrm{coarse}}$ (Line~8--9 of Algorithm~\ref{alg:coarsening-padding}).
    Note that unlike Algorithm~\ref{alg:original-padding}, there is no need to perform FFTs on an enlarged FFT grid as padding the coarsened cell returns the original FFT grid.
    \item Subtract from $V_{\mathrm{periodic}}$ the unwanted periodic interactions $V_{\mathrm{periodic,coarse}}$ and add the correctly truncated interactions $V_{\mathrm{coarse}}$ to generate the potential under the required boundary conditions (Line~10--11 of Algorithm~\ref{alg:coarsening-padding}).
\end{enumerate}
The reasoning behind the present approach is loosely based on the Gaussian theorem, {\em i.e.} long-range electrostatics depend only on monopoles, dipoles, and higher-order poles. It is well understood that broadening the density, e.g. replacing a point charge with a Gaussian charge density, leaves the electrostatic potential unchanged beyond the radius where the Gaussian decays to zero.\cite{jackson_classical_1999} Removing large $\bf g$ components from the density also preserves all electrostatic multipoles, thus completely preserving the long-range electrostatics. This means that the current approach is inherently free of any errors, which we will demonstrate below.
In passing, we note that the FFT grids used to correct the errors in the long-range electrostatics could be further reduced, but from an implementation point of view, the present strategy was easier to implement in VASP, and a further reduction would only marginally increase the computational efficiency. 

\subsection{No-padding method}

An alternative approach to truncating the Coulomb kernel is to avoid padding entirely\cite{Sohier2017,Rozzi}.
This no-padding method implicitly assumes that the charge density decays to near zero at the boundary of the cell along the non-periodic dimension.
This assumption allows for truncation without padding with the explicit consideration that the potential generated at the edges of the cell are non-physical and are to be ignored (dark regions of Fig.\ \ref{fig:coarsen-schematic}c).
This method is implemented by simply swapping out $v(\mathbf{g})$ for the appropriately truncated kernel.

\subsection{Comparison of self-consistent potentials}

\begin{figure}[!htb]
    \centering
    \includegraphics[width=\linewidth]{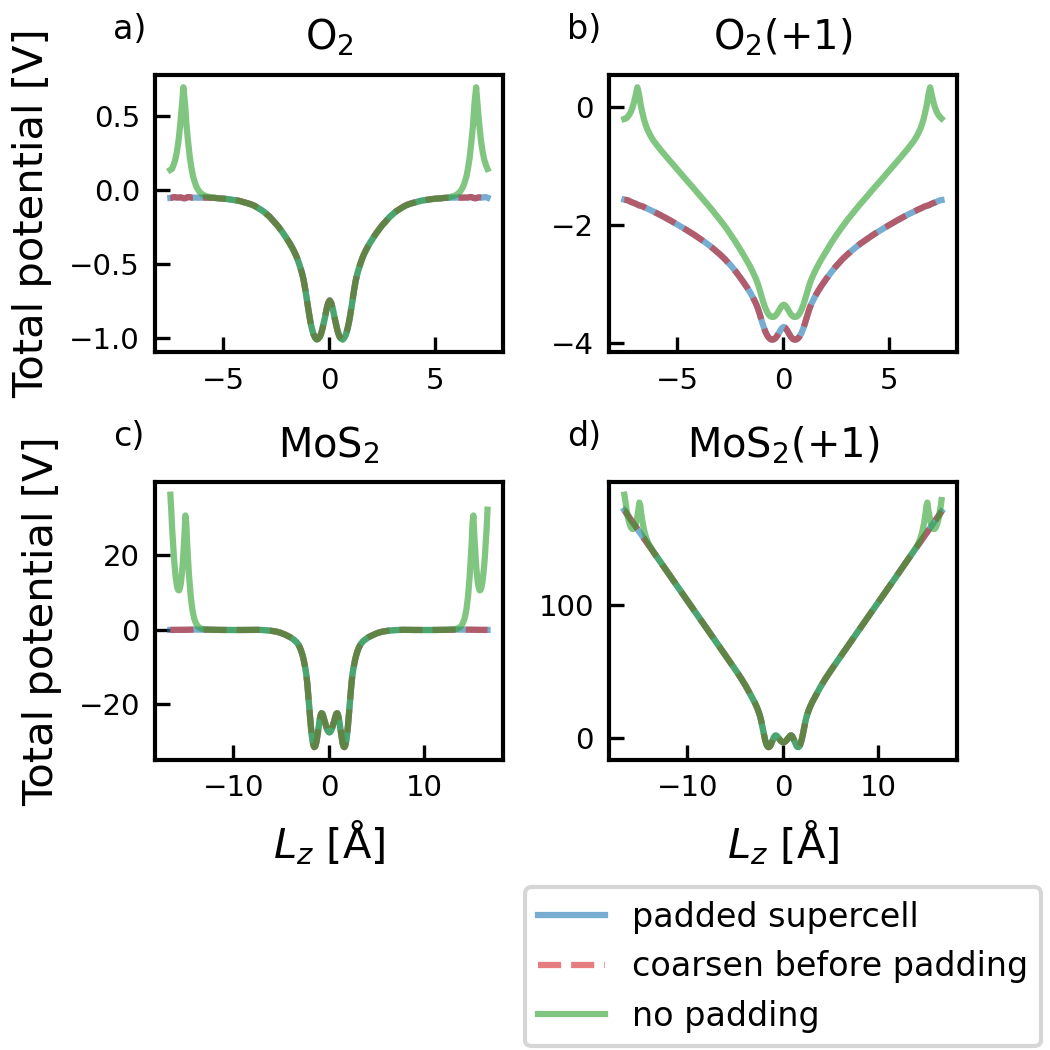}
    \caption{Total potential generated by the padded supercell method (solid blue line), the coarsen before padding method (dashed red line), and the no-padding method (solid green line) a) oxygen molecule b) positively charged oxygen molecule c) a single layer of MoS$_2$ d) a positively charged single layer of MoS$_2$.}
    \label{fig:compare-truncate-potential}
\end{figure}

Fig.\ \ref{fig:compare-truncate-potential} shows that the total potential computed using the coarsen before padding method (dashed red lines) matches the total potential computed by the conventional padded supercell method (solid blue lines) for a charge neutral and charged O$_2$ molecule and a single layer of MoS$_2$.
This potential equivalence shows that we can achieve exactly the same accuracy for our DFT calculation without the need for expensive FFT operations on a larger FFT grid.

Both the coarsen before padding and the padded supercell method generate physical potentials throughout the cell.
In contrast, the no-padding approach (green lines, Fig.\ \ref{fig:compare-truncate-potential}) produces physical potentials when the charge density is non-zero (for example, at $L_z \approx 0$), but produces non-physical potentials towards the cell boundaries ($L_z =\pm 6$~\AA\ for O$_2$ and $L_z = \pm 12$~\AA\ for MoS$_2$).

Note that the potentials of the no-padding approach are similar to the padded supercell and coarsen before padding potentials for the case of MoS$_2$(+1), but appear slightly different for O$_2$(+1).
This deviation comes from the choice of $R_\mathrm{c}$ used for the no-padding approach.
We set $R_\mathrm{c}$ to half the length of the largest norm of the lattice vectors (padded or otherwise) for MoS$_2$ and choose $R_\mathrm{c}$ as $\sqrt{3}$ times the cell for O$_2$ computed with the padded supercell method and coarsen before padding method. 
For O$_2$ computed with the no-padding approach, we set $R_\mathrm{c}$ as half the length of the largest norm of the lattice vectors.
This choice is required for the no-padding approach as it does not pad the cell and hence $R_\mathrm{c}$  cannot be larger than 1.

Well-chosen $R_\mathrm{c}$ ensures that the non-physical region is distinct from any charge density, preventing any unwanted interactions from taking place.
However, this convergence with respect to $R_\mathrm{c}$ may not be possible, especially during molecular dynamics calculations involving ``floppy" molecules which may move into regions of unphysical potential.

Given the excellent performance of our coarsen before padding method at a reduced computational cost as compared with the padded supercell method and the need for a physical potential throughout the cell, we set it as the default option for VASP calculations including Coulomb kernel truncation and use it for the rest of this manuscript.
\section{Energies for 0D and 2D boundary conditions}
\label{sec:energies}

In Sections~\ref{sec:boundary_conditions} and \ref{sec:coarsening}, we showed that by using an appropriate Coulomb truncated kernel, we generate the Hartree, ionic and total potentials for 0D and 2D boundary conditions.
In this section, we briefly describe how these potentials are used to compute components of the total energy.
We highlight modifications to VASP to facilitate total energy calculations with 0D and 2D boundary conditions.

\subsection{Hartree energy}
The electrostatic energy caused by electron-electron interactions (Hartree energy) is computed in reciprocal space as
\begin{equation}
    \label{eq:hartree-energy}
    E_{\mathrm{Hartree}} = \frac{\Omega}{2} \sum_{\mathbf{g}} \rho(\mathbf{g}) V(\mathbf{g}),
\end{equation}
where $V(\mathbf{g})$ is the potential under either 3D, 2D or 0D boundary conditions as determined in Sections \ref{sec:boundary_conditions} and \ref{sec:coarsening}.
The only change to the current VASP implementation is to supply $V(\mathbf{g})$ for the appropriate boundary condition to Eq.\ (\ref{eq:hartree-energy}).

\subsection{Sum of eigenvalues}

The sum of the eigenvalues on a per-atom basis is
\begin{equation}
    \label{eq:sum-eigenvalues}
    E_{\mathrm{eigenvalues}} = \sum_{i}\epsilon_i,
\end{equation}
\noindent where $i$ is the running index over all computed eigenvalues with a non-zero occupancy.
Eq.\ (\ref{eq:sum-eigenvalues}) implicitly depends on the total potential as the numerical values of $\epsilon$ are determined by the potential in which they are computed. 

For 3D boundary conditions, the value of the potential is arbitrary (as the $\mathrm{g=0}$ component of Hartree and ionic potentials are set to zero) and hence the the values of $\epsilon$ are arbitrary as well.
In contrast, for both 0D and 2D boundary conditions, we include the $\mathrm{g}=0$ component (as seen in Eqs.\ (\ref{eq:kernel-0D}) and (\ref{eq:kernel-2D}) respectively) explicitly. This implementation choice allows us to compute the absolute eigenvalues for charge-neutral molecules using the 0D boundary condition and surfaces using the 2D boundary condition.
This output of the absolute eigenvalues implies that properties such as the workfunction of 2D surfaces are immediately available and do not require a potential reference within the cell.

\subsection{Ion-ion interation energy}

As done for 3D boundary conditions implementation in VASP, we compute the ion-ion interaction energies using the Ewald summation technique. 
The total summation of ion point charges on a lattice is split into a real space (short-range) component and a reciprocal space (long-range) component with some additional analytical terms
\begin{equation}
\begin{split}
    \label{eq:ewald-sum}
    E_{\mathrm{ion-ion}} =  & E_{\mathrm{long-range}} + E_{\mathrm{short-range}} + E_{\mathrm{self}} + \\
    & + E_{\mathrm{homogeneous}},
    \end{split}
\end{equation}
where $E_{\mathrm{long-range}}$ is the long-range energy, $E_{\mathrm{short-range}}$ is the short-range energy, $E_{\mathrm{self}}$ is the energy of self-interaction between the smeared charge and the point charge of the Ewald sum and $E_{\mathrm{homogeneous}}$ is the energy of interaction between the point charge and a homogeneous background charge.

For 0D boundary conditions, we set $E_{\mathrm{short-range}}$, $E_{\mathrm{self}}$ and $E_{\mathrm{homogeneous}}$ to zero and compute the entire ion-ion interaction energy in real space. We choose this direct computational strategy as there is no periodicity and hence no need for long-range interaction terms,  self correction and homogeneous background charge terms,
\begin{equation}
    E_{\mathrm{ion-ion}} = \frac{1}{2}\sum_{i\neq j} \frac{Z_iZ_j}{r_{ij}},
\end{equation}
where $r_{ij}$ is the distance between any two atomic charges with index $i$ and $j$ and $Z_i$ and $Z_j$ are the corresponding atomic charges.

For 2D boundary conditions, we implement the 2D Ewald summation to compute the terms in Eq.~\ref{eq:ewald-sum} as per the formulation of Ref.~\onlinecite{heyes_molecular_1977}, which includes terms for $E_{\mathrm{long-range}}$, $E_{\mathrm{short-range}}$ and $E_{\mathrm{self}}$, while settings $E_{\mathrm{homogeneous}}$ to zero.
The long-range part is
\begin{widetext}
\begin{equation}
\begin{split}
    E_{\mathrm{long-range}} &= \frac{\pi}{2A}\sum_{i,j} Z_i Z_j \sum_{\mathrm{g}\neq 0} \cos(\mathbf{g} \cdot \mathbf{r}_{ij}) \\ &\left [ \frac{ \exp(\mathrm{g}z) \mathrm{erfc}\left(\alpha z + \mathrm{g}/(2\alpha) \right) + \exp(-\mathrm{g}z) \mathrm{erfc}\left(-\alpha z + \mathrm{g}/(2\alpha) \right)}{\mathrm{g}} \right],
\end{split}
\end{equation}
\end{widetext}
where $A$ is the surface area, $z$ is the distance between $i$ and $j$ in the direction of the surface normal and $\alpha$ is the splitting parameter between long and short-range summations.
We do not change the default value of $\alpha$ used by VASP, which is inversely proportional to the cube root of the volume of the cell.
The short-range energy is computed similarly as done with 3D boundary conditions,
\begin{equation}
    E_{\mathrm{short-range}} = \sum_{i,j}^\prime  Z_i Z_j \sum_{\mathbf{n}} \frac{\mathrm{erfc}\left( \alpha \left | \mathbf{r}_{ij} + \mathbf{n} \right | \right)}{\left | \mathbf{r}_{ij} + \mathbf{n} \right |},
\end{equation}
where $\mathbf{n}$ is the index of the real space cell and the prime over the first summation indicates that the sum does not include terms where $i=j$ and $\mathbf{n}=0$.
The self-correction term is
\begin{widetext}
\begin{equation}
    E_{\mathrm{self}} = -\frac{\alpha}{\sqrt{\pi}} \sum_{i=1}^{N}Z_i^2 -\frac{2\pi}{A} \sum_{i,j} Z_i Z_j  \left[ z\mathrm{erf}(\alpha z) + \frac{\exp\left(-(z\alpha)^2\right)}{\alpha\sqrt{\pi}} \right],
\end{equation}
\end{widetext}
where $N$ is the number of atoms.
The term in square brackets is an additional self-correction term for systems with a nonzero height parallel to the surface normal or systems that are not charge neutral.

\subsection{Validation}

We test our implementation by computing the self-consistent Hartree energy, eigenvalues, ion-ion energies and total energies for a charge neutral pyridene molecule under 0D boundary conditions and charged single layer MoS$_2$ sheet under 2D boundary conditions.
We describe the behavior of these energies with change in lattice vectors along all directions for pyridene and along the surface normal (i.e.\ amount of vacuum) for MoS$_2$.

\begin{figure*}
    \centering
    \includegraphics[width=0.6\linewidth]{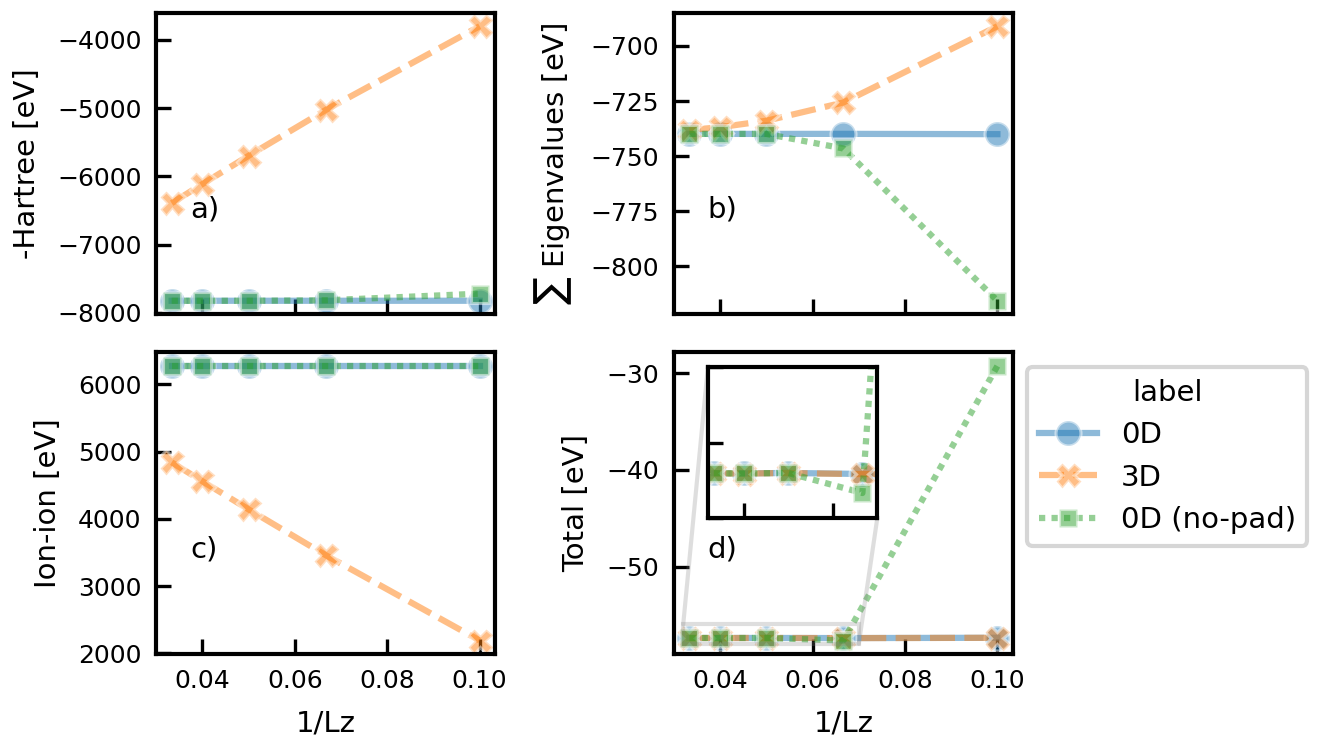}
    \caption{
    Variation of energy contributions for a charge-neutral pyridine molecule: a) negative of the Hartree energy, b) sum of the eigenvalues, c) ion-ion energy, and d) total energy; orange points and lines indicate that 3D boundary conditions were used, blue points and lines indicate 0D boundary conditions were used, and green points and lines indicate that 0D boundary conditions using the no-padding method were used. The inset shows a zoom-in of the total energies.}
    \label{fig:comparison-energy-pyridene}
\end{figure*}

\subsubsection{Charge neutral pyridine molecule}

Figure \ref{fig:comparison-energy-pyridene} shows the (negative of) the Hartree energy, sum of the eigenvalues, ion-ion and total energies for pyridine under 3D boundary conditions (in orange), 2D boundary conditions (in blue) and 2D boundary conditions with no-padding (in green) with change in the vacuum in the cell ($x$-axis is the inverse of the vacuum lattice dimension, $z$).
Applying 0D boundary conditions leads to energies that are invariant to the amount of vacuum in the cell, i.e.\ all blue points are equivalent.
In contrast, Hartree energies, eigenvalues, and the ion-ion energies computed by the 3D boundary conditions are dependent on the choice of vacuum.
This dependence on the vacuum size exactly cancels for total energies of charge-neutral materials (Fig.~\ref{fig:comparison-energy-pyridene}d).
This match of the total energy between the 0D and 3D boundary conditions (blue and orange points overlap in Fig.~\ref{fig:comparison-energy-pyridene}d) shows that our implementation yields identical results to the 3D implementation when sufficient vacuum is used for charge-neutral species.

Finally, we compare the total energies of the coarsen before padding and no-padding method.
We find that the coarsen before pad method converges with less vacuum in the cell than the no-padding method (total energies at $1/L_z = 0.1 \mathrm{\AA}^{-1}$ in Fig.\ \ref{fig:comparison-energy-pyridene}~d).
This faster convergence with vacuum dimension implies that the coarsen before padding method can be used with less vacuum in the cell for 0D systems.

\subsubsection{Positively charged MoS$_2$ sheet}

\begin{figure*}
    \centering
    \includegraphics[width=0.6\linewidth]{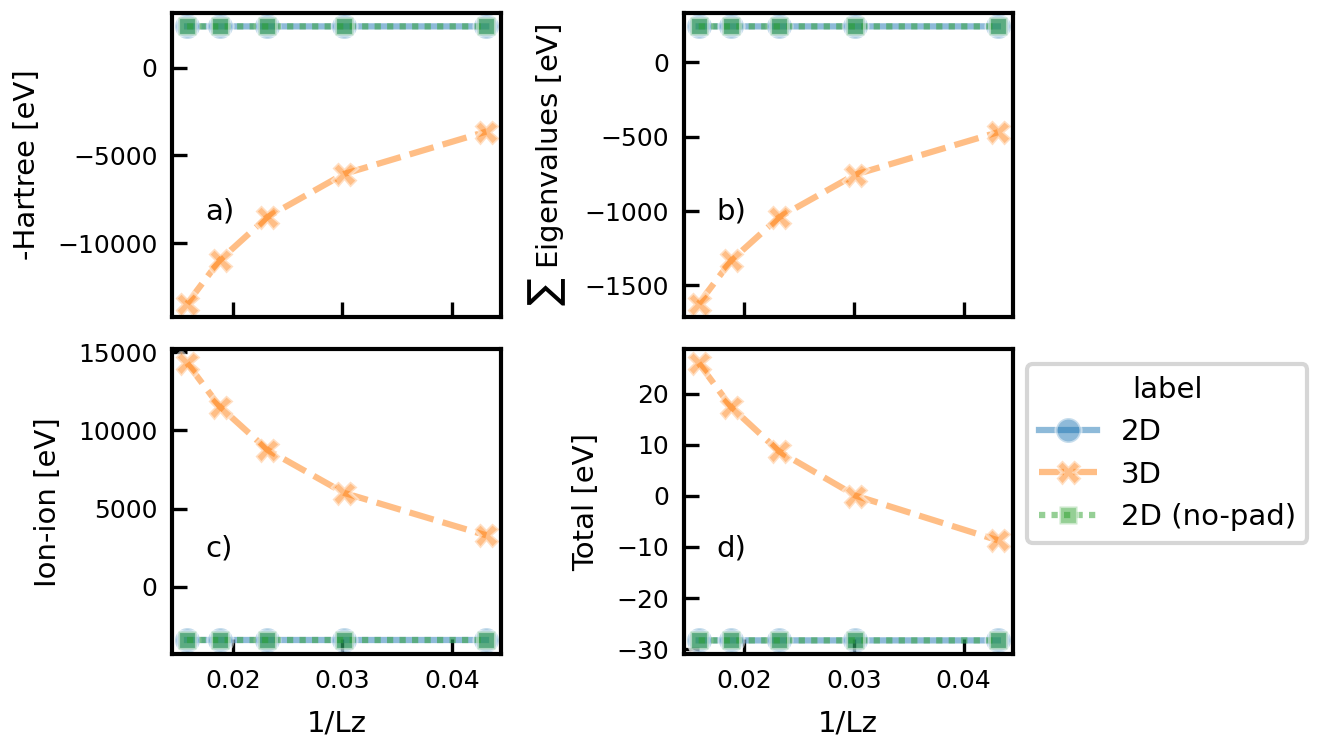}
    \caption{
    Variation of energy contributions for a charged MoS$_2$ sheet: a) negative of the Hartree energy, b) sum of the eigenvalues, c) ion-ion energy, and d) total energy; orange points and lines indicate 3D boundary conditions were used, blue points and lines indicate 2D boundary conditions were used, and green points and lines indicate that 2D boundary conditions using the no-padding method were used.} 
    \label{fig:comparison-energy-mos2}
\end{figure*}

A key difference between energies computed with 3D and 2D boundary conditions is the convergence behavior of the total energies for positively charged MoS$_2$ (in Figure \ref{fig:comparison-energy-mos2}).
The total energy (Figure \ref{fig:comparison-energy-mos2}d) is identical for all vacuum sizes when computed using 2D boundary conditions (blue and green points). 
In contrast, the total energies diverge when computed with 3D boundary conditions (orange points in Figure \ref{fig:comparison-energy-mos2}d), i.e.\ they increase with increasing vacuum size and never converge to a single value.
This divergence of the total energy stems from the inability of 3D boundary conditions to balance the individual divergences of the Hartree energy (orange points, Figure \ref{fig:comparison-energy-mos2}a), eigenvalues (orange points, Figure \ref{fig:comparison-energy-mos2}b) and ion-ion energy (orange points, Figure \ref{fig:comparison-energy-mos2}c). In comparison, the Hartree energies, sum of eigenvalues, ion-ion and total energies computed with 2D boundary conditions show no divergence (blue and green points in Figure \ref{fig:comparison-energy-mos2}~a,b,c).

We note that the success of the 2D method to provide consistent energies is in part related to the choice that the potentials at the center of the slabs converge to one value independent of the vacuum width, whereas the potential at large distances diverges (as $L_z$).
This behavior of the potential means that the work for adding or removing a charge quanta from a slab is always infinite, consistent with simple electrostatic arguments,\cite{jackson_classical_1999} and the divergence of the energies for the 3D case.
The cost to generate a 2D charged slab is always infinite, but it is still possible to compare the stability of different charged 2D structures.
This comparison is simplified using our implementation of 2D boundary conditions.

In summary, we find that self-consistent calculations with 0D and 2D boundary conditions for charge neutral and charged surfaces provide consistent energies.
This behavior is in contrast to performing calculations with 3D boundary conditions, which lead to convergent energies for charge-neutral surfaces but divergent energies for charged molecules and surfaces.

\section{Applications}
\label{sec:applications}

In this section, we apply our implementation of the Coulomb kernel truncation method to two prototype cases of charged DFT calculations. 
Specifically, we compute the concentration dependent formation energy of a chlorine defect on NaCl(001) and the total energies and potential profile from a long-timescale machine-learning assisted Au(211) $\vline$ water interface.
Both illustrations would be particularly expensive with the padded supercell method due to either a requirement for large cells or long time scale simulations.

\subsection{Formation energy of a charged defect on a surface}

\begin{figure*}[!htb]
    \centering
    \includegraphics[width=0.8\linewidth]{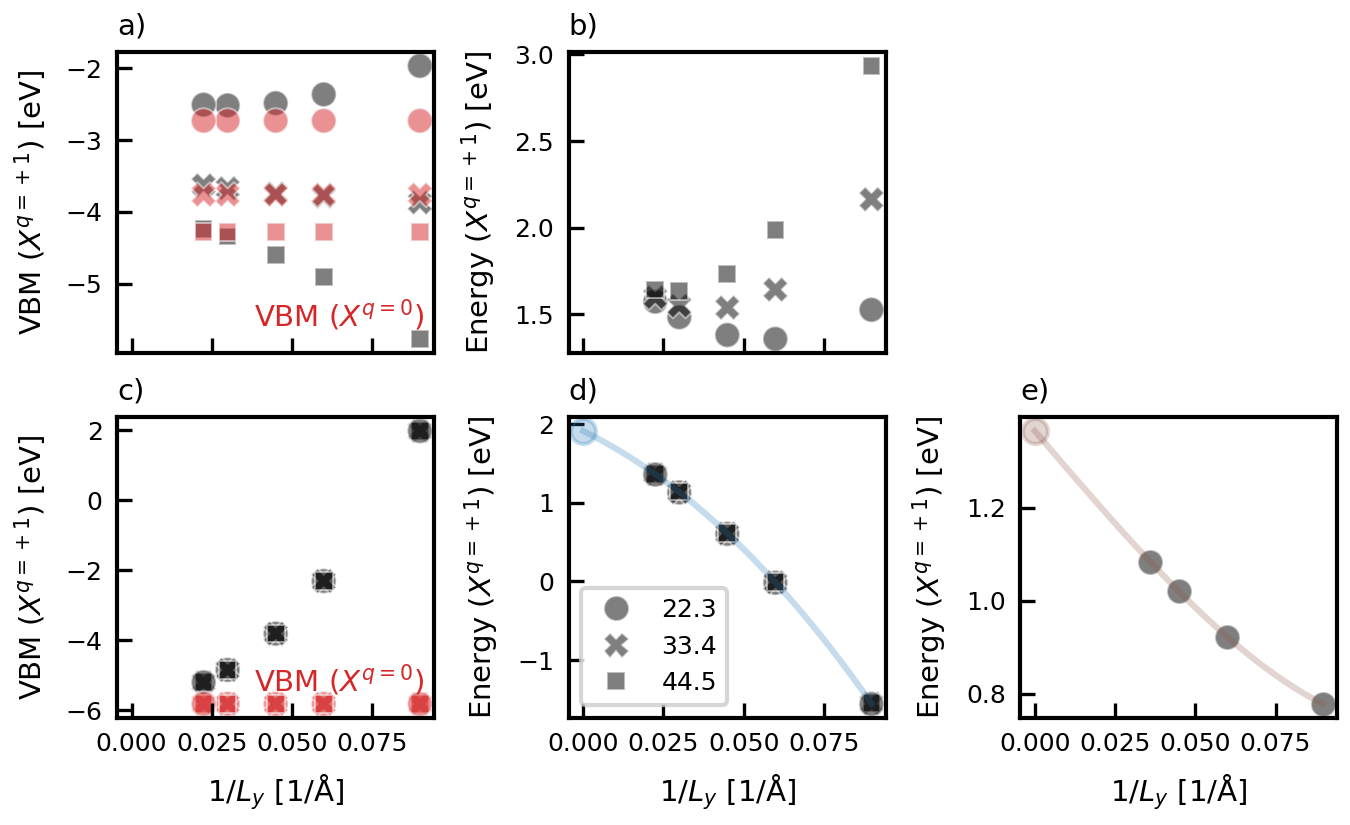}
    \caption{Valence band maximum (VBM) computed under a) 3D and d) 2D boundary conditions. Grey markers denote the VBM of the NaCl Cl-vacancy charged defect and red points correspond to the VBM of the uncharged pristine surface. Formation energies of the charged defect calculated under b) 3D and d) 2D boundary conditions. There is no dependence on the vacuum dimension with 2D boundary conditions (overlapping markers). Blue solid line in (d) shows the least squares fit for a cubic function. e) Formation energy of a charged Cl-vacancy defect in bulk NaCl; brown solid line shows the least squares fit for a function of $1/L_y+1/L_y^3$. Root mean square errors for the least square fits are 0.004~eV for surface defects (blue line in d) and 0.0001~eV for bulk defects (brown line in e). An equivalent quadratic fit for 3D  formation energies does not produce a good fit with an error of 0.29~eV. The extrapolated value of $1/L_y\to 0$ is shown as a blue dot in (d) and a brown dot in (e). Markers indicate changing vacuum size.}
    \label{fig:charged-defect}
\end{figure*}

To illustrate the difference between 3D and 2D boundary conditions, we compute the concentration-dependent formation energy of a prototypical charged defect, a chlorine vacancy defect on a surface of NaCl(001).

Formation energies of charged defects on surfaces show strong dependence on the concentration of the defects and vacuum when computed with 3D boundary conditions.\cite{Freysoldt2009,Freysoldt2014,Komsa2013,Komsa2014}
Several methods have been proposed to remove this dependence on the concentration and vacuum.\cite{Komsa2013,Freysoldt2009}
In general, these methods generate a surrogate model (typically a Gaussian charge density around the defect) in successively larger cells and extrapolate the energy to the limit of infinitely large vacuum and infinitely small defect concentration.\cite{Freysoldt2009}

In this section, we show that the concentration and vacuum-dependent formation energy for charged defects computed under 2D boundary conditions show comparatively simple behavior, leading to a simpler extrapolation scheme using a quadratic least squares fit.

The formation energy of a charged defect of Cl in NaCl(001) is,
\begin{widetext}
\begin{equation}
    E_f[\ce{Na}_n\ce{Cl}_{n-1}(+1)] = E[\ce{Na}_n\ce{Cl}_{n-1}(+1)] - E[\ce{Na}_n\ce{Cl}_n] + \frac{1}{2} E[\ce{Cl2(g)}] + q\left(\epsilon_{v} + \epsilon_{F} \right),
\end{equation}
\end{widetext}
where $E_f$ is the formation energy, $E[\ce{Na}_n\ce{Cl}_{n-1}(+1)]$ is the total energy of the charged defect with $n$ Na species and $n-1$ Cl species, $E[\ce{Na}_n\ce{Cl}_n]$ is the total energy of the pristine host (without the defect) for $n$ Na and Cl species, $E[\ce{Cl2(g)}] $ is the chemical potential for the missing chlorine species between the host and the defect, $\epsilon_F$ is the Fermi energy which is referenced to the valence band maximum (VBM) $\epsilon_{v}$.
As we are interested in the general behavior of the formation energy with concentration and vacuum, we set $\epsilon_F=0$.
Any other value of $\epsilon_{F}$ would be a constant offset of the energies reported in this section.

Fig.\ \ref{fig:charged-defect} illustrates differences in the behavior of the VBM  and formation energies of the charged defect (denoted as $X^{q=1}$) as a function of the defect concentration (by changing cell dimension, denoted as $1/L_y$) and vacuum size (markers) computed under 3D and 2D boundary conditions.
We modify the defect concentration by constructing a single chlorine defect within an increasingly large cell size (\textit{x}-axis in Fig.\ \ref{fig:charged-defect}) and alter the vacuum by changing the cell dimension along the surface normal (markers in Fig.\ \ref{fig:charged-defect}).
Fig.\ \ref{fig:charged-defect}a shows the VBM of a surface charged defect (grey points) and of the pristine surface (red points, denoted as $X^{q=0})$ as a function of decreasing surface concentration of the defect (\textit{x}-axis is the inverse of the lattice dimension along the surface, $1/L_y$) computed with 3D boundary conditions. 
Consistent with the arbitrary offset of the potential with 3D boundary conditions, we find that the VBM is different for different surface concentrations (\textit{x}-axis) and different vacuum sizes (markers).
There is no clearly extrapolatable trend as a function of different cell dimensions.
Fig.\ \ref{fig:charged-defect}c shows the VBM computed with 2D boundary conditions. 
In contrast to Fig.\ \ref{fig:charged-defect}a, we find that the VBM has no dependence on vacuum size (overlapping grey points) and decays to the VBM of the uncharged surface (red points).
This behavior of the VBM is consistent with the reduced concentration of the charge as $1/L_y$ goes to zero.
In this case, using 2D boundary conditions leads to extrapolatable and intuitive behavior of the VBM of charged defects.

Formation energies do not converge with cell size (Fig.~\ref{fig:charged-defect}b) when computed under 3D boundary conditions.
As a result, there is no clear trend between the formation energy and the defect concentration.
\textit{Post facto} methods, which treat the interaction of the charged defect with different concentrations of the homogeneous background charge\cite{Komsa2013,Komsa2014,Freysoldt2009} are needed to determine the formation energy.

In contrast, formation energies computed with 2D boundary conditions converge with decreasing surface concentration (Fig.~\ref{fig:charged-defect}d).
Furthermore, there is no dependence of the formation energies on the vacuum (markers in Fig.\ \ref{fig:charged-defect}d).
This comparatively straightforward behavior suggests that extrapolation techniques can be used to determine the formation energy of the charged defect under infinitely dilute concentration.

To validate our approach of computing formation energies using 2D boundary conditions, we extrapolate the energies to $1/L_y \to 0$ and compare our results against previously published work. \cite{Komsa2013}
We fit the formation energies shown in Figure \ref{fig:charged-defect}d (grey points) by least squares fitting a third-order polynomial and extrapolate it to $1/L_y\to 0$ (blue curve in Figure \ref{fig:charged-defect}d).
Our rationale for choosing a linear, quadratic and cubic terms is to effectively account for monopole-monopole, monopole-dipole and dipole-dipole interactions, which are likely to be the leading terms in the dependence of the energy on the change in the defect concentration.
We note that fitting up to the quadratic term (i.e.\ ignoring the cubic term) results in less than 0.01~eV change in the extrapolated formation energies for this system.
We find the formation energy at infinitely dilute concentrations to be 1.91~eV, which is very similar to the value of 1.89~eV computed by extrapolating formation energies using 3D boundary conditions and a model charge distribution.\cite{Komsa2013}
The root mean square error (RMSE) of the fit is 0.004~eV, suggesting that our effective model of monopole-monopole and dipole-dipole interactions are sufficient to describe the interactions between charged defects on a surface.

Figure \ref{fig:charged-defect}e shows a fit including linear and cubic terms, i.e.\ $1/L_y + 1/L_y^3$ (brown line) for a charged Cl-vacancy defect in bulk NaCl.
We compute its formation energy under 3D boundary conditions, identical to the method used in the case of the charged surface defect.
Similar to the surface defect in Figure \ref{fig:charged-defect}d, we expect the leading concentration-dependent terms to be monopole-monopole and dipole-dipole terms; the monopole-dipole terms are zero for NaCl as it is an isotropic system.
We find that the least squares error for this fit is 0.0001~eV, suggesting a good fit between this simple model and the computed formation energies.
Extrapolating this fit to $1/L_y\to 0$ leads to a formation energy at infinitely dilute concentrations of 1.365~eV, which is similar to the value of 1.50~eV obtained using a model charge distribution.\cite{Komsa2013}

In summary, the formation energy of a surface-charged defect of chlorine on an NaCl(001) surface displays contrasting behavior with decreasing defect concentration and vacuum size for 3D and 2D boundary conditions.
For 3D boundary conditions, both the VBM and the formation energies do not converge and require special methods to compute both of these quantities in the infinitely dilute concentration limit.
In contrast, both the VBM and the formation energies display convergent behavior with calculations performed with 2D boundary conditions.
This convergent behavior simplifies extrapolating formation energies to the infinite dilute concentration limit.

\subsection{Electrode electrolyte interface}
\begin{figure*}[!htb]
    \centering
    \includegraphics[width=0.7\linewidth]{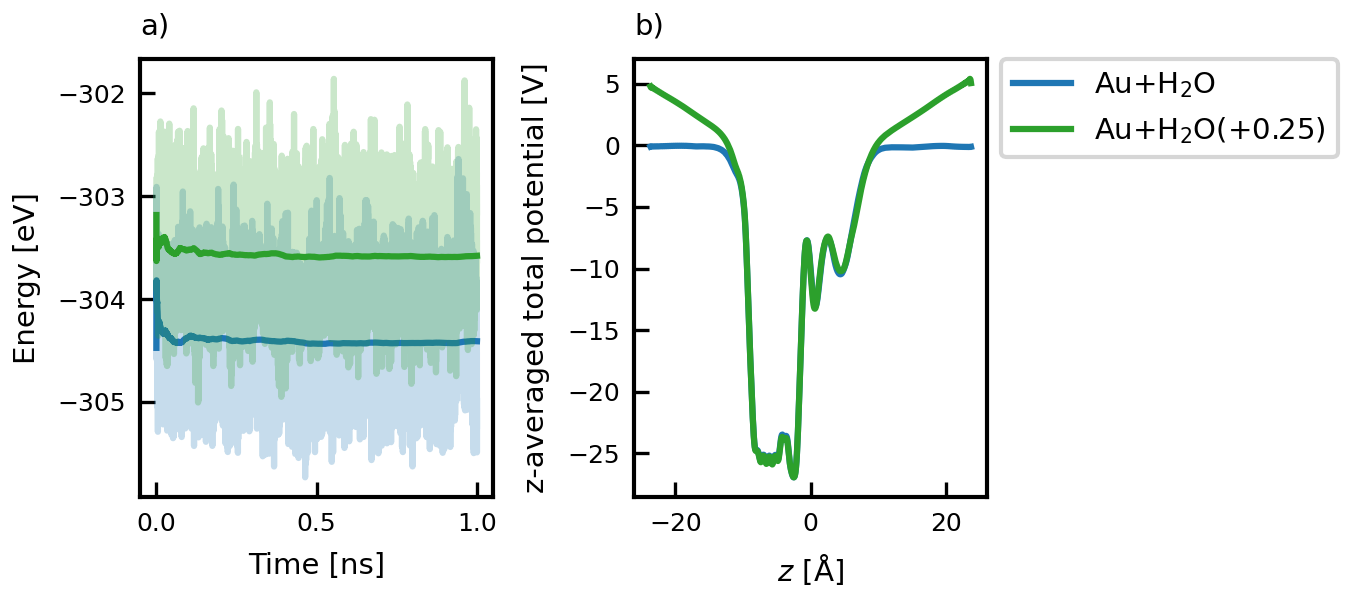}
    \caption{a) Total energy for each time step (transparent lines) and cumulative average (solid lines) for a (blue) charge neutral and (green) positively charged Au(211) $\vline$ water interface b) \textit{xy}-averaged total potential for 100 evenly spaced snapshots from a 1ns trajectory. The potential for the charged surface (green lines) has been moved to overlap with that of the uncharged (blue) electrostatic potential.}
    \label{fig:charged-electrolyte}
\end{figure*}

A key advantage of the coarsen before padding method is that the computational cost is roughly the same as for the equivalent 3D boundary condition calculation.
This advantage allows us to perform \textit{ab initio} molecular dynamics calculations for charged systems with 2D boundary conditions.
To illustrate this advantage, we train a machine-learned force field (see Section \ref{sec:methods}) using more than 2400 DFT calculations with 2D boundary conditions for a system of 15 water molecules interacting with an Au(211) surface. We train two distinct force fields, one that is charge neutral and one with an added excess charge of 0.25~e.

Fig.\ \ref{fig:charged-electrolyte}a shows the cumulative average energies (solid lines) and instantaneous energies (translucent lines) of a 1 nano-second (ns) NVT molecular dynamics calculation for uncharged (blue) and charged (green) water \vline \ Au(211) interface.
The energies of both trajectories are stable for both charge-neutral and charged systems, i.e.\ the cumulative energies do not fluctuate.
The charge-neutral system has a lower total energy due to the presence of more electrons and differences in the computed potential.

Fig.\ \ref{fig:charged-electrolyte}b shows the \textit{z}-averaged electrostatic potential for 100 evenly spaced structures from the 1~ns molecular dynamics trajectory.
The charge neutral potential (blue) decays to zero in the vacuum region ($z\approx 20~\text{\AA}$).
This flattening of the electrostatic potential to zero suggests that the averaged orientation of the water molecules is arbitrary, {\em i.e.} the dipoles of water do not specifically orient along a certain axis.
In cases where water had a specific mean orientation, we would expect the averaged electrostatic potential to decay to a nonzero value, with the potential at both ends ($z=\pm 10~\mathrm{\AA}$) settling at different values.

The \textit{z}-averaged electrostatic potential increases linearly for a charged simulation cell (green lines in Fig.\ \ref{fig:charged-electrolyte}b).
This behavior of the potential for the charged system is similar to that of the charged plate potential of Figure \ref{fig:comparison-potential} and is a consistent feature of applying 2D boundary conditions to charged systems, irrespective of the thickness of the material.\cite{jackson_classical_1999}

Comparison of the electrostatic potentials of the charged and uncharged slab in Fig.\ \ref{fig:charged-electrolyte}b shows that the addition of a small amount of charge creates a field that is completely shielded by the aqueous electrolyte layer: 
the electrostatic potentials with and without charge completely overlap in the region of the aqueous electrolyte (0 -- 10 $\mathrm{\AA}$), suggesting that the linear potential slope (visible between 10 -- 20 $\mathrm{\AA}$) is entirely screened. In other words, the mesoscopic $E$ field $E=D-4 \pi P$ is practically zero in the water layer.
This screening is in line with the large dielectric constant of bulk water of 78.4 and implies that in the water layer the polarization observes almost exactly $P=D/4 \pi$.
This result also means that the water molecules reorient compared to the uncharged field-free case. A more detailed analysis is beyond the scope of the present work, albeit there is clearly interesting physics to unravel with the present approach.

In summary, we applied the Coulomb kernel truncation method to study the formation energy of charged defects and performed long time-scale molecular dynamics simulations of water over a charge-neutral and charged Au(211) surface.
We find that our implementation provides a natural basis in which to compute properties of charged surfaces. 

Given the convergent behavior for charged systems and the comparatively low computational cost of our implementation, we envision this Coulomb kernel truncation implementation to provide a starting point for further method development for computing properties of charged molecules and surfaces.
\section{Conclusions}

In this work, we describe our Coulomb kernel truncation implementation in the VASP package.
Our implementation allows to perform DFT calculations under 0D and 2D boundary conditions for charge-neutral and charged systems.
Instead of solving the Poisson equation to determine the potential on a computationally expensive vacuum-padded supercell, our approach subtracts only the relevant long-range contributions from the potential obtained in 3D boundary conditions.
By using this method of determining the potential, our approach is only marginally more computationally expensive than calculations using 3D boundary conditions. The present approach could easily be adopted for calculations of the exact exchange potential in Hartree-Fock or hybrid functional calculations, where we expect it to have an even greater impact than in the present case. We note that we have been rather modest in the grid coarsening in the present work, as for density functional theory the calculation of the potential is usually not a dramatic bottleneck. However, the coarsening can be increased significantly without loss of accuracy, since only electrostatic multipoles need to be described on the coarse grid.

We have applied our method to calculate the formation energy of a charged chlorine defect on a NaCl(001) surface.
We find that the VBM and formation energy diverge when computed with 3D boundary conditions, but converge when computed with 2D boundary conditions. This allows reliable Cl vacancy formation energies to be extrapolated from the slab calculations. The extrapolated formation energy shows very good agreement with previous calculations.
Furthermore, we demonstrate the computational efficiency of our implementation by performing nanosecond timescale molecular dynamics simulations of the Au(211) $\vline$ water interface, with an excess charge located on the Au slab.  This approach allows us to generate the long-range electrostatic field required to mimic an electrode. We believe that the present approach is well suited to determine structural changes induced by the finite field at an electrode. In fact, we observe that the water molecules rearrange in order to fully screen the displacement field caused by the Au-electrode. 
\section{Methods}
\label{sec:methods}

All density functional theory calculations were performed with the Vienna \textit{ab-initio} Simulation Package (VASP).\cite{Kresse1996}
The Coulomb kernel truncation method is implemented in VASP and is available as of version 6.5.0. 

Charged defect calculations were performed on a single chlorine defect of NaCl(001).
The structure, cell and volume of bulk NaCl were relaxed using the \verb|ISIF=3| option implemented in VASP using the PBE functional with a \textit{k}-point sampling of 12$\times$12$\times$12, yielding a lattice constant of $a=3.93~\mathrm{\AA}$, $c=5.56~\mathrm{\AA}$.
The \verb|Na_pv| and \verb|Cl| PAW potential were used for Na and Cl atoms respectively.
A plane-wave cutoff of 262~eV was used for all charged defect calculations.
The \textit{k}-point density was kept constant for different defect concentrations on NaCl(001) by dividing the number of repetitions of the cell by 12, i.e.\ a 6$\times$6 \textit{xy}--repetition of the cell was computed with a 2$\times$2$\times$1 \textit{k}-point sampling.
Gaussian smearing was used with a broadening on 0.1~eV for all calculations.
Similar to Ref.~\onlinecite{Komsa2013}, no surface relaxation was performed for the defect as we are interested in visualizing qualitative differences between 3D and 2D boundary conditions.

Molecular dynamics simulations of a 3$\times$3 Au(211) surface with 4 Au layers in contact with an interface of water (3 layers with a total of 15 water molecules) were performed with an NVT ensemble.
The lattice constant of Au ($a=4.17~\mathrm{\AA}$) was determined based on the same computational setup as used for the molecular dynamics simulations.
The \verb|Au|, \verb|O| and \verb|H| PAW potentials were used for Au, O and H species respectively.
The bottom two layers were fixed in position throughout the simulation to mimic bulk-Au.
The temperature was controlled using the Nose-Hoover\cite{Nose1984} thermostat.
The mass of hydrogen in all simulations was set to 8 au.

We train a machine-learned force field (kernel-based method implemented in VASP)\cite{jinnouchi_phase_2019} to generate long time scale simulation trajectories.
Training structures to the force field were generated on-the-fly using the RPBE\cite{Hammer1999} functional with D3\cite{Grimme2006} dispersion correction and zero damping.
The temperature for both training runs was set to 300K for the first 1800 training structures and then set to 350K for the remaining structures.
The training error for charge-neutral Au(211) $\vline$ water is $1.15$~meV/atom for the total energies and $77.9$~meV/\AA{} for the forces with $2445$ DFT computations.
The training error for positively charged (with a charge of 0.25 electrons removed) Au(211) $\vline$ water interface is $1.27$~meV/atom for the total energies and $76.0$~meV/\AA{} with $2822$ DFT computations. 

Post-processing of all potential profiles used in this manuscript was performed with \verb|py4vasp| and visualization was performed with \verb|matplotlib|.
\section*{Author contribution}

\textbf{Sudarshan Vijay}: Conceptualization, Data curation, Formal analysis, Investigation, Software, Methodology, Validation, Visualization, Writing – original draft, Writing – review \& editing. 
\textbf{Martin Schlipf}: Software, Visualization, Writing – review \& editing.
\textbf{Henrique Miranda}: Formal analysis, Writing – review \& editing.
\textbf{Ferenc Karsai}: Formal analysis, Writing – review \& editing.
\textbf{Merzuk Kaltak}: Formal analysis, Software, Writing -review \& editing.
\textbf{Martijn Marsman}: Formal analysis, Software, Writing – review \& editing.
\textbf{Georg Kresse}: Conceptualization, Funding acquisition, Investigation, Methodology, Software, Resources, Writing – review \& editing
\section*{Acknowledgement}

S.V., M.S., M.M. and G.K. acknowledge support form EuroHPC Joint Undertaking for awarding us access to Leonardo at CINECA, Italy. This research was funded in part by the Austrian Science Fund (FWF) 10.55776/COE5. For open access purposes, the author has applied a CC BY public copyright license to any author accepted manuscript version arising from this submission.  
\section*{Conflict of interest}

All authors of this manuscript are fully or partially employed by the VASP Software GmbH and are developers of the Vienna \textit{ab-initio} Software Package.

\bibliography{references}

\end{document}